\documentclass[aps,prd,onecolumn,nofootinbib]{revtex4}
\usepackage{epsfig}
\usepackage{graphicx}
\usepackage{dcolumn}
\usepackage{amsmath}
\usepackage{enumerate}
\usepackage{amssymb}
\usepackage{latexsym}
\usepackage{bbold}
\usepackage{bbm}
\usepackage[mathscr]{eucal}
\usepackage[english]{babel}
\def\a{\alpha}

\def\d{\delta}

\def\l{\lambda}
\def\m{\mu}

\def\t{\tau}

\def\y{\eta}

\def\f{\varphi}

\def\half{\frac{1}{2}}

\def\mtil{\widetilde{m}}
\usepackage{color}
\definecolor{rosso}{cmyk}{0,1,1,0.4}
\definecolor{rossos}{cmyk}{0,1,1,0.55}
\definecolor{rossoc}{cmyk}{0,1,1,0.2}
\definecolor{blu}{cmyk}{1,1,0,0.3}
\definecolor{blus}{cmyk}{1,1,0,0.6}
\definecolor{bluc}{cmyk}{1,1,0,0.1}
\definecolor{verde}{cmyk}{0.92,0,0.59,0.25}
\definecolor{verdec}{cmyk}{0.92,0,0.59,0.15}
\definecolor{verdes}{cmyk}{0.92,0,0.59,0.4}

\def\circa#1{\,\raise.3ex\hbox{$#1$\kern-.75em\lower1ex\hbox{$\sim$}}\,}

\newcommand{\eV}{\,{\rm eV}}

\newcommand{\GeV}{\,{\rm GeV}}

\newcommand{\beq}{\begin{equation}}
\newcommand{\eeq}{\end{equation}}
\newcommand{\bea}{\begin{eqnarray}}
\newcommand{\eea}{\end{eqnarray}}
\newcommand{\ba}{\begin{array}}
\newcommand{\ea}{\end{array}}
\newcommand{\eps}{\epsilon}

\newcommand{\gsim}{\lower.7ex\hbox{$\;\stackrel{\textstyle>}{\sim}\;$}}
\newcommand{\lsim}{\lower.7ex\hbox{$\;\stackrel{\textstyle<}{\sim}\;$}}

\newcommand{\baz}{\begin{array}{cc}}
\newcommand{\bad}{\begin{array}{ccc}}
\def\gtap{\mathrel{ \rlap{\raise 0.511ex \hbox{$>$}}{\lower 0.511ex
   \hbox{$\sim$}}}} 
\def\ltap{\mathrel{ \rlap{\raise 0.511ex
   \hbox{$<$}}{\lower 0.511ex \hbox{$\sim$}}}}
   \newcommand{\deltaatm}{\mbox{$\Delta m^2_{\mathrm{A}}$}}
   \newcommand{\deltasol}{\mbox{$ \Delta m^2_{\odot}$}}
\newcommand{\dmsol}{\mbox{$\Delta m^2_{\odot}$}}
\newcommand{\dma}{\mbox{$\Delta m^2_{\rm A}$}}

\newcommand{\pmns}{\mbox{$ U$}}

\begin{document}

\begin{flushright}
SISSA 52/2008/EP\\
\end{flushright}


\title{
A Case of Subdominant/Suppressed ``High Energy'' 
Contribution to the Baryon Asymmetry 
of the Universe in Flavoured Leptogenesis}

\author{E.~Molinaro$^{a)}$, S.~T.~Petcov$^{a,b)}$
\footnote{Also at: Institute
of Nuclear Research and Nuclear Energy, Bulgarian Academy 
of Sciences, 1784 Sofia, Bulgaria.}
}

\affiliation{{\it $^{a)}$SISSA and INFN-Sezione di Trieste, 
Trieste I-34014, Italy}}
\affiliation{{\it $^{b)}$IPMU, University of Tokyo, Tokyo, Japan}}


\begin{abstract}

The CP-violation necessary for the generation 
of the baryon asymmetry of the Universe $Y_B$
in the ``flavoured'' leptogenesis scenario
can arise from the ``low energy'' PMNS 
neutrino mixing matrix $U$ and/or from the 
``high energy'' part of neutrino Yukawa couplings, 
which can mediate CP-violating phenomena 
only at some high energy scale.
The possible interplay between these two types of 
CP-violation is analysed. The type I see-saw model 
with three heavy right-handed Majorana neutrinos 
having hierarchical spectrum is considered.  
We show that in the case of inverted 
hierarchical light neutrino mass spectrum,
there exist regions in the corresponding 
leptogenesis parameter space 
where the relevant ``high energy'' 
phases have large CP-violating
values, but the purely ``high energy''
contribution in $Y_B$ plays a subdominant role 
in the production of baryon asymmetry 
compatible with the observations.
In some of these regions the purely
``high energy'' contribution in $Y_B$ 
is so strongly suppressed that 
one can have successful leptogenesis 
only if the requisite CP-violation is provided 
by the Majorana phase(s) in the neutrino 
mixing matrix.

\end{abstract}

\maketitle
It is well established at present 
\cite{davidsonetal,davidsonetal2}
(see also \cite{Barbieri99,Nielsen02}) 
that lepton flavour effects can play a 
very important role in the leptogenesis 
mechanism \cite{FY,kuzmin} of generation 
of the baryon asymmetry of the
Universe, $Y_B$. In the regime in which the 
lepton flavour effects in leptogenesis 
are significant (``flavoured'' leptogenesis),
the CP-violation necessary for the generation 
of matter-antimatter 
asymmetry compatible with the
observations can be provided exclusively by 
\cite{PPRio106} 
the Dirac and/or Majorana \cite{BHP80} 
CP-violating phases in the 
Pontecorvo-Maki-Nakagawa-Sakata (PMNS)
neutrino mixing matrix 
$U_{\rm PMNS} \equiv U$ \cite{BPont57}. 
In the case of three hierarchical 
heavy right-handed (RH) Majorana 
neutrinos $N_j$, $j=1,2,3$, 
and CP violation due to the Majorana phases 
in $U$, this typically requires  
that the mass of the lightest RH Majorana neutrino 
$N_1$, satisfies 
$M_1 \gtap 4\times 10^{10}$ GeV \cite{PPRio106}.
In ref. \cite{MPST07}
it was shown that if the requisite CP-violation 
in ``flavoured'' leptogenesis is 
due to the CP violating phases in the 
PMNS matrix, 
the baryon asymmetry $Y_B$ in certain
physically interesting cases 
exhibits strong dependence 
on the lightest neutrino mass, 
${\rm min}(m_j)$, $j=1,2,3$.
For specific values of 
${\rm min}(m_j)$, in particular,
the asymmetry $Y_B$ can be strongly enhanced 
(by a factor of $\sim 100$ or more) 
with respect to that predicted 
in the case of ${\rm min}(m_j) =0$.
This enhancement can make the predicted $Y_B$ 
compatible with the observations even when 
this is not the case for ${\rm min}(m_j) \cong 0$.
Some aspects of the matter-antimatter asymmetry 
generation in the ``flavoured'' leptogenesis
scenario in the case when the relevant CP-violation 
is due to the Majorana or Dirac CP-violating 
phases in $U_{\rm PMNS}$, were investigated in 
refs. \cite{SBDibari06,DiBGRaff06}.

   In this work we analyse 
the more general 
possibility in which
the requisite CP-violation in ``flavoured'' 
leptogenesis is provided both by the ``low energy'' 
Majorana and/or Dirac CP-violating phases 
in the neutrino mixing matrix 
and the ``high energy'' phases
which can be present in the matrix 
of neutrino Yukawa coupling, $\l$, 
and can mediate CP-violating processes 
only at some ``high'' energy scale. 
The scheme in which we work is the 
non-supersymmetric type I see-saw model \cite{seesaw} 
with three heavy right-handed (RH) 
Majorana neutrinos, $N_j$, having masses $M_j$ with
hierarchical spectrum, $M_1\ll M_{2} \ll M_{3} $. 
The see-saw mechanism of neutrino mass generation 
provides a natural explanation of the smallness of 
neutrino masses. Moreover, through the leptogenesis 
theory it allows to relate the generation and 
the smallness of neutrino masses with the generation 
of the baryon asymmetry of the Universe, $Y_B$. 
In the thermal leptogenesis scenario, 
the CP-violating asymmetry
relevant for leptogenesis in the case of hierarchical 
heavy Majorana neutrino masses, is generated in 
out-of-equilibrium decays of the lightest RH 
neutrino, $N_1$. The latter is produced by thermal 
scattering after inflation.

  In the basis in which the Majorana mass matrix of the 
RH neutrinos and the matrix of the charged 
lepton Yukawa couplings, $\l^{lep}$, are diagonal, 
the only source of CP-violation in the lepton sector 
is the matrix of neutrino Yukawa couplings $\l$. 
The orthogonal parametrization
of $\l$ \cite{Casas01}, involving a 
complex orthogonal matrix $R$, 
allows to relate in a simple way $\l$ 
with the neutrino mixing matrix $U$: 
$\lambda = (1/v)\sqrt{M} \, R\, \sqrt{m} \, U^{\dagger}$, 
where $M$ and $m$ are diagonal matrices formed by
the masses $M_j > 0$ and $m_k \geq 0$ of $N_j$ and 
of the light Majorana neutrinos $\nu_k$ respectively, 
$j,k=1,2,3$, and $v = 174$ GeV is the vacuum expectation 
value of the Higgs doublet field. This parametrization
permits to  investigate the combined effect 
of the CP-violation due to the ``low energy'' 
neutrino mixing matrix $U$ 
and the CP-violation due to the ``high energy'' 
matrix $R$ in the generation of the baryon asymmetry in 
``flavoured'' leptogenesis. 
The PMNS matrix $U$ is present 
in the weak charged lepton current 
and can be a source of CP-violation 
in, e.g. neutrino oscillations at 
``low'' energies
\footnote{As is well-known, 
only the Dirac phase in $U$ can be a source of 
CP-violation in neutrino oscillations;
the probabilities of oscillations of
flavour neutrinos do not depend on the 
Majorana phases in $U$ \cite{BHP80,Lang87}.}
$E \sim M_Z$ 
(see, e.g. \cite{BiPet87,PKSP3nu88,Future}). 
The matrix $R$, as is 
well-known, does not affect the ``low'' 
energy neutrino mixing phenomenology.
The two matrices $U$ and $R$
are, in general, independent. 
It should be noted, however, 
that in certain specific cases 
(of, e.g. symmetries and/or texture zeros)
of the matrix $\lambda$ of neutrino Yukawa couplings, 
there can exist a relation 
between (some of) the CP-violating phases in 
$U$ and (some of) the CP-violating 
parameters in $R$ (see, e.g. \cite{PRST05,PSMoriond06}).
For hierarchical heavy Majorana neutrinos, 
the baryon asymmetry $Y_B$ depends on the 
CP-violating (complex) elements $R_{1j}$, $j=1,2,3$, 
of the $R-$matrix.

  In the present letter we study certain 
aspects of the possible interplay between 
the ``low energy'' CP-violation 
due to the Dirac and/or Majorana CP-violating 
phases in the PMNS matrix $U$, 
and the ``high energy'' CP-violation
originating from the matrix $R$, 
in ``flavoured'' leptogenesis.  
We concentrate on the case of
light Majorana neutrinos with 
inverted hierarchical (IH) spectrum 
(see, e.g. \cite{STPNu04}), 
$m_3\,\ll\,m_{1,2}\,\cong\,\sqrt{|\dma|}\,\cong\,0.05\,\eV$.
The case of normal hierarchical 
spectrum has been analysed in detail in 
~\cite{Molinaro:2008rg}, 
were results for the IH spectrum 
were also presented. Here we  
investigate in greater detail
the case of IH spectrum,
thus extending further the analysis
performed in \cite{Molinaro:2008rg}.
We show that if the light 
Majorana neutrinos  
possess IH mass spectrum,
there exist significant regions of the 
corresponding leptogenesis 
parameter space where the relevant 
``high energy'' $R-$phases have 
large CP-violating values, but
the purely ``high energy'' 
contribution in $Y_B$ 
plays a subdominant role 
in the production of 
baryon asymmetry compatible with 
the observations. 
The requisite dominant term in $Y_B$ 
can arise due to 
the ``low energy'' CP-violation in 
the neutrino mixing matrix $U$. 
In some of these regions 
the ``high energy'' contribution in $Y_B$ 
is so strongly suppressed 
that one can have successful leptogenesis 
only if the requisite CP-violation is provided 
by the Majorana CP-violating phase(s) in $U$.

     Negative results regarding the possible 
effects of ``low energy'' CP violation in 
``flavoured'' leptogenesis with
hierarchical heavy (RH) and light Majorana 
neutrinos, when the  ``high energy'' 
CP-violation is also present,
were reported in \cite{SDavid07} and more 
recently in \cite{SDavid08}.
We would like to note that the study 
performed in \cite{SDavid07} was by no 
means exhaustive. More specifically,
the results reported in \cite{SDavid07} were obtained
either for i) ${\rm min}(m_j) =0$ and specific
texture zero in the $R$-matrix,
or  ii) for a value of the mass of the
lightest RH neutrino
\footnote{Private communication by S. Davidson. 
We thank S. Davidson for clarifications 
regarding the analysis 
performed in \cite{SDavid07}.}
$M_1 = 10^{10}$ GeV.
In the case ii), the lightest neutrino mass
${\rm min}(m_j)$ was allowed to vary within
the interval $0 \leq {\rm min}(m_j) \leq 10^{-3}$ eV.
However, in both these cases the contribution in 
$Y_B$ due to the ``low energy'' CP violation
is strongly suppressed. The contribution 
under discussion can
be relevant for the production of $Y_B$ 
compatible with the observations 
provided \cite{PPRio106}
$M_1 \gtap 4\times 10^{10}$ GeV;  
in the case i) it can be relevant 
if one considers values of
$M_1 \gtap 4\times 10^{10}$ GeV and of
${\rm min}(m_j) \gtap 5\times 10^{-4}$ eV \cite{MPST07}.  
In the present article we explore the region 
of the parameter space corresponding to 
$M_1 \gtap 5\times 10^{10}$ GeV. In this region 
the effects of the ``low energy'' 
CP violation in flavoured leptogenesis 
can be significant. In what concerns the analysis 
performed in \cite{SDavid08}, it differs 
substantially from the analysis performed here.
In \cite{SDavid08} the leptogenesis is 
considered in the framework of the 
SUSY extension of the Standard Model, 
more specifically, in the 
minimal Supergravity (MSUGRA)
scenario with real boundary conditions, 
in which the dynamics responsible for supersymmetry
breaking are flavour blind and all the lepton flavour and 
CP violation is controlled by the neutrino Yukawa couplings. 
The leptogenesis parameter space 
is constrained, in particular,
by requiring that the $\mu \rightarrow e + \gamma$ 
decay rate branching ratio, predicted in this scenario,
satisfies $BR(\mu \rightarrow e + \gamma) \geq 10^{-12}$. 
We work in the simpler non-SUSY version 
of leptogenesis. The difference between our results and 
those found in \cite{SDavid08} 
may reflect the difference in the 
priors on the scanned leptogenesis parameters, 
for instance, the range in which 
the lightest RH neutrino mass $M_1$ is varied.

  We use the standard parametrization 
of the PMNS neutrino mixing matrix:
\begin{widetext}
\bea 
\label{eq:Upara}
\pmns = \left( \bad 
c_{12} c_{13} & s_{12} c_{13} & s_{13}e^{-i \delta}  \\[0.2cm] 
 -s_{12} c_{23} - c_{12} s_{23} s_{13} e^{i \delta} 
 & c_{12} c_{23} - s_{12} s_{23} s_{13} e^{i \delta} 
& s_{23} c_{13}  \\[0.2cm] 
 s_{12} s_{23} - c_{12} c_{23} s_{13} e^{i \delta} & 
 - c_{12} s_{23} - s_{12} c_{23} s_{13} e^{i \delta} 
 & c_{23} c_{13} \\ 
                \ea   \right) 
~{\rm diag}(1, e^{i \frac{\alpha_{21}}{2}}, e^{i \frac{\alpha_{31}}{2}})
\eea
\end{widetext}
%
\noindent where $c_{ij} \equiv \cos\theta_{ij}$, $s_{ij} \equiv
\sin\theta_{ij}$, $\theta_{ij} = [0,\pi/2]$, $\delta = [0,2\pi]$ is
the Dirac CP-violating phase and $\alpha_{21}$ and $\alpha_{31}$
are the two Majorana CPV phases \cite{BHP80,SchValle80Doi81},
$\alpha_{21,31} = [0,4\pi]$. All our numerical results 
are obtained for the best fit values of the solar 
and atmospheric neutrino oscillation parameters 
\cite{BCGPRKL2,Fogli06,TSchw08}: 
$\deltasol=\Delta m^2_{21} = 7.65 \times 10^{-5}\,\eV^2$,
$\sin^2 \theta_{12} = 0.30$,
$|\deltaatm| = |\Delta m^2_{31(32)}| = 
2.4 \times 10^{-3}\,\eV^2$ and $\sin^2 2\theta_{23} = 1$. 
We also use the upper limit on the CHOOZ mixing angle
$\theta_{13}$ 
\cite{CHOOZ,BCGPRKL2,TSchw08}:
$\sin^2\theta_{13} < 0.035~(0.056)\,,~~
95\%~(99.73\%)~{\rm C.L.}$. 

   We work in the ``two flavour'' regime
~\cite{davidsonetal,davidsonetal2},
in which the $\t$ flavour interactions are 
in thermal equilibrium and the Boltzmann evolution of
the CP-asymmetry in the $\tau$ lepton charge, 
$\eps_\t$, is distinguishable from 
the evolution of the ($e+\m$)-flavour asymmetry
$\eps_2\equiv\eps_e+\eps_\m$. This regime is realised 
at temperatures $10^9\,\GeV\ltap T\sim M_1\ltap 10^{12}\,\GeV$. 
In this study
we neglect the effects of 
the lightest neutrino mass $m_3$ 
and we set for simplicity $|R_{13}|=0$. 
The latter condition 
is compatible with the hypothesis of  
$N_3$ decoupling in the case of IH spectrum 
\footnote{A complete decoupling of $N_3$ 
in the case of IH spectrum
occurs when the elements of the R-matrix 
satisfy $R_{13} = R_{23} = R_{31} = R_{32} = 0$ 
\cite{PRST05}.
In the context of ``flavoured'' 
leptogenesis the case of $N_3$ decoupling 
and $|R_{13}|=0$ for the IH spectrum
was discussed earlier in 
\cite{davidsonetal2,PPRio106}. 
However, our analysis 
practically does not overlap
with the analyses performed in 
\cite{davidsonetal2,PPRio106}. 
Let us note finally that the $N_3$ 
decoupling in the case 
of NH spectrum was considered, e.g. in 
\cite{FGY03,IR041,PRST05}.}
\cite{PRST05}. 
The results of our 
analysis are valid actually
if the following much 
weaker conditions are fulfilled:
if $|R_{13}|^2 |\sin2\tilde{\f}_{13}| \ll 
{\rm min} (|R_{11}|^2 |\sin2\tilde{\f}_{11}|,
|R_{12}|^2 |\sin2\tilde{\f}_{12}|)$, 
and if $m_3$  is sufficiently small, so that 
the terms $\propto m_3 |R_{13}|^2$ and 
$\propto m^2_3 |R_{13}|^2$ in the asymmetries 
$\epsilon_2$ and $\epsilon_{\tau}$ 
are negligible. The first condition
will be satisfied even if $R^2_{13}$ 
is not zero, but is just real,
i.e. if ${\rm Im}(R^2_{13}) = 0$.
The second condition is 
naturally satisfied in the case 
of IH spectrum.

    From the orthogonality condition for the 
$R-$matrix elements of interest in the general case of
$R_{13} \neq 0$, ${\rm Im}(R^2_{13}) = 0$,
$R_{11}^2 + R_{12}^2 + R^2_{13}=1$, 
$R_{1j}\equiv|R_{1j}|e^{i\tilde{\f}_{1j}},\,j=1,2$,
we can express the ``high energy'' CP-violating phases, 
$\tilde{\f}_{11}$ and $\tilde{\f}_{12}$, in terms
of the absolute values $|R_{11}|$, $|R_{12}|$ and of $R^2_{13}$ 
which is real: 
\begin{eqnarray}
\cos 2\tilde{\f}_{11} & = &
\frac{\left (1 - R^2_{13}\right)^2 + |R_{11}|^4 - |R_{12}|^4}
{2|R_{11}|^2\left(1 - R^2_{13}\right)}\,,
\label{tef11} \\
\cos 2\tilde{\f}_{12} & = &
\frac{\left (1 - R^2_{13}\right)^2 - |R_{11}|^4 + |R_{12}|^4  }
{2|R_{12}|^2\left(1 - R^2_{13}\right)}\,,~~
\label{tef12} 
\end{eqnarray}
%
with ${\rm sgn}(\sin 2\tilde{\f}_{11}) = -{\rm sgn}(\sin 2\tilde{\f}_{12})$.
In the cases we discuss below (and illustrate in Figs. 1 and 2) 
we have chosen $\sin 2\tilde{\f}_{11} < 0$. 
The CP-violating asymmetry $\eps_\t$ in
the case considered is given by:
\begin{widetext}
\begin{eqnarray}\label{CP-asym-epstau_IH}
\eps_\t & \cong & 
-\frac{3\,M_1}{16\,\pi\, v^2}\,\frac{\sqrt{|\dma|}}{|R_{11}|^2\,
+\,|R_{12}|^2}\,\left\{|R_{11}|^2\,\sin(2\tilde{\f}_{11})\,
\left[ (|U_{\t1}|^2 - |U_{\t2}|^2) -  
 \frac{\dmsol}{|\dma|}\,|U_{\t1}|^2\right]\right.\nonumber\\\\
& + &
\left.|R_{11}|\,|R_{12}|\left[\half\,\frac{\dmsol}{|\dma|}\,
\cos(\tilde{\f}_{11}+\tilde{\f}_{12})\,{\rm Im}(U_{\t1}^*U_{\t2})\,
+ \,2\,\left(1 - \half\,\frac{\dmsol}{|\dma|}\right)\,
\sin(\tilde{\f}_{11}+\tilde{\f}_{12})
\,{\rm Re}(U_{\t1}^*U_{\t2})\right]\right\}\nonumber
\end{eqnarray}
\end{widetext}
For $\tilde{\f}_{11}=k\pi/2$,
$\tilde{\f}_{12}=k'\pi/2$, $k,k'=0,1,2,...$,
$R_{11}$ and $R_{12}$ are either real or purely 
imaginary and the expression for $\eps_\t$ 
reduces to the one derived in \cite{PPRio106}. 
Under these conditions we can have successful 
leptogenesis for $R_{13} = 0$ 
in the case considered only 
if $R_{11}R_{12}$ is purely imaginary, 
i.e. if $|\sin(\tilde{\f}_{11}+\tilde{\f}_{12})|=1$,
the requisite CP-violation being provided 
exclusively by the Majorana or Dirac phases in 
the PMNS matrix \cite{PPRio106}. 
We remind the reader that 
i) the $R$-matrix will satisfy the CP-invariance 
constraint if its elements $R_{ij}$ 
are real or purely imaginary 
\footnote{For the precise form of the 
CP-invariance constraint on the elements 
$R_{ij}$ of the $R$-matrix see 
\cite{PPRio106}.}, and 
ii)  in order to have CP-violation,
e.g. only due to the Majorana phase
$\alpha_{21}$ in $U$, 
both  ${\rm Im}(U_{\t1}^*U_{\t2})$
and ${\rm Re}(U_{\t1}^*U_{\t2})$ should be 
different from zero \cite{JMaj87,ASBranco00},
while the Dirac phase $\delta$ should have 
a CP-conserving value, $\delta = k\pi$, $k=0,1,2,...$
(i.e., the rephasing invariant $J_{\rm CP}$ associated 
with $\delta$ \cite{PKSP3nu88} should satisfy $J_{\rm CP} = 0$).
Let us note also that purely imaginary
$R_{11}R_{12}$, i.e. $|\sin(\tilde{\f}_{11}+\tilde{\f}_{12})|=1$,
and  ${\rm Re}(U_{\t1}^*U_{\t2}) = 0$, $J_{\rm CP} = 0$
corresponds to the case of CP-invariance and
$\eps_\t = 0$. 
However, purely imaginary $R_{11}R_{12}$ and  
$J_{\rm CP} = 0$, ${\rm Im}(U_{\t1}^*U_{\t2}) = 0$, 
but ${\rm Re}(U_{\t1}^*U_{\t2}) \neq 0$ 
(i.e. $\delta = k\pi$,  $\alpha_{21} = 2\pi q$,
$k,q=0,1,2,...$), corresponds to 
CP-violation due to the neutrino Yukawa couplings, i.e.
due to the combined effect of the matrix $R$ 
and of the PMNS matrix $U$ \cite{PPRio106},  
and $\eps_\t \neq 0$. It is interesting that 
in this case both the $R$-matrix and 
the PMNS matrix $U$ satisfy the 
CP-invariance constraints 
(having real and/or purely imaginary elements), 
while the neutrino Yukawa couplings do not
satisfy these constraints. As a consequence, 
under the indicated conditions
i) there will be no 
CP-violation effects caused by PMNS matrix $U$
in the low energy neutrino mixing phenomena 
(neutrino oscillations, neutrinoless double beta decay, etc.),
and ii) there will be no CP-violation effects in the 
``high energy'' phenomena which depend only on 
the matrix $R$ (i.e. do not depend on the 
PMNS matrix $U$). 

  Indeed, consider the general case of 
hierarchical heavy RH Majorana neutrinos
with arbitrary light neutrino mass spectrum,
non-negligible lightest neutrino mass and 
matrix $R$ with non-zero elements. 
The CP-violating asymmetry 
\begin{figure}[t!!]
\begin{center}
\includegraphics[width=8.5cm,height=6.5cm]{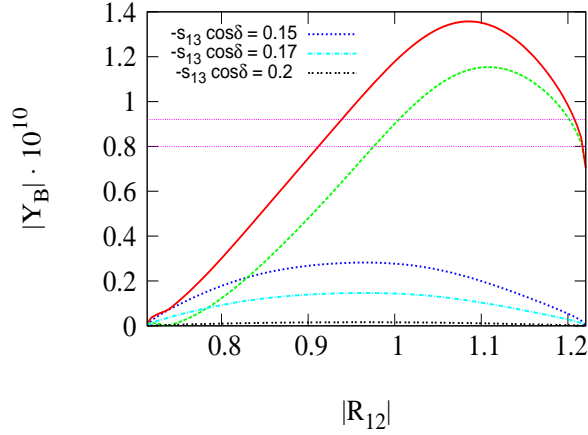}
\caption{
\label{IH_YB_a21_0.5a} 
The dependence of 
the ``high energy'' term 
$|Y^0_B A_{\rm HE}|$ (blue line), the ``mixed'' term
$|Y^0_B A_{\rm MIX}|$ (green line) and of the 
total baryon asymmetry $|Y_B|$ (red line) 
on $|R_{12}|$ in the case of IH spectrum, 
CP-violation due to the Majorana phase
$\a_{21}$
and $R$-phases, for $(-s_{13}\,\cos\d)=0.15;~0.17;~0.20$, 
$\alpha_{21} = \pi/2$, $|R_{11}| = 0.7$,
$M_1 = 10^{11}$ GeV. 
The horizontal lines indicate the allowed range of 
$|Y_B|$, $|Y_B|=(8.0-9.2)\times 10^{-11}$.
}
\end{center}
\end{figure}
%
\noindent in the lepton flavour $l$, generated in the 
decays of the lightest RH Majorana 
neutrino $N_1$ reads (see, e.g. \cite{PPRio106}):
\begin{equation}
\epsilon_{l} = -\frac{3 M_1}{16\pi v^2}\, \frac{{\rm Im}\left(
\sum_{j,k} m_j^{1/2}m_k^{3/2} U^*_{l j}U_{l k}
R_{1j}R_{1k}\right)}{\sum_\beta m_\beta\left|R_{1\beta}\right|^2}\,,~~
l=e,\mu,\tau\,.
\label{epsl}
\end{equation}
%
Let us assume next that $M_1 > 10^{12}$ GeV.
In this case the baryon asymmetry $Y_B$ is 
determined by the sum of the CP-violating asymmetries 
in the individual lepton flavours 
\cite{LG1,LG2,davidsonetal}:
\begin{equation}
\epsilon_{1} = \sum_{l}{\epsilon_{l}} 
= -\frac{3 M_1}{16\pi v^2}\, \frac{{\rm Im}\left(
\sum_{j} m_j^{2} R^2_{1j}\right)}
{\sum_k m_k\left|R_{1k}\right|^2}\,,~~
l=e,\mu,\tau\,.
\label{eps1}
\end{equation}
%
As it follows from the preceding equation and
is well-known, the asymmetry $\epsilon_{1}$
does not depend on the PMNS matrix - it depends 
only on the elements of the $R$-matrix. 
In this sense one can say that  
the generation of the baryon asymmetry  
in the regime under discussion 
is purely a ``high energy'' phenomenon.
It should be obvious, however, from eq. (\ref{eps1}) 
that if the $R$-matrix is CP-conserving, i.e. 
if its elements $R_{1j}$ are real 
\cite{davidsonetal} (E. Nardi et al.) 
or purely imaginary \cite{PPRio106}, we would have 
\footnote{As can be shown, 
this result does not depend, 
in particular, on the assumption  
about the spectrum of the heavy 
Majorana neutrinos. More concretely,
the CP violating asymmetry $\epsilon_{j}$, 
generated in the decays of the heavy Majorana 
neutrino $N_j$ due to the interference of 
the tree level and one loop contributions 
to the relevant amplitudes, would be 
zero if  $M_j > 10^{12}$ GeV, $j=1,2,3$,
and $R_{jk}$ is either real or purely 
imaginary, $k=1,2,3$ and satisfies the 
CP-invariance constraint \cite{PPRio106}.}
$\epsilon_{1} = 0$.
In what concerns the ``low energy'' 
neutrino mixing phenomena caused by the 
PMNS matrix $U$, the CP-symmetry 
will obviously be unbroken for  
$\delta = k\pi$,  $\alpha_{21} = 2\pi q$ and
$\alpha_{31} = \pi q'$, $k,q,q'=0,1,2,...$.

 One can easily show that for 
the IH light neutrino mass spectrum
of interest in the present analysis, 
the following relation holds 
\footnote{Note that this relation is valid 
not only for $R_{13}=0$, but also for
nonzero real $R^2_{13}$, $R_{13} \neq 0$,
${\rm Im}(R^2_{13}) = 0$.}
: $\eps_2 = -\eps_\t(1 + O(\dmsol/|\dma|))$. 
Thus, the baryon asymmetry $Y_B$ can be written 
as a function of $\eps_\t$ only, like in the 
case of the matrix $R$ satisfying the CP-invariance 
constraints \cite{PPRio106}:
\begin{eqnarray}
Y_B & = &-\frac{12}{37}\frac{\eps_\t}{g_*}
\left(\, \y\left(\frac{390}{589}\mtil_{\t}\right)\, 
- \,\y\left(\frac{417}{589}\mtil_2\right)\right)\nonumber\\
    &\equiv& Y^0_B(A_{\rm HE}+A_{\rm MIX})\label{YB_2}
\end{eqnarray}
\begin{figure}[t!!]
\begin{center}
\begin{tabular}{c}
 \includegraphics[width=8.5cm,height=6.5cm]{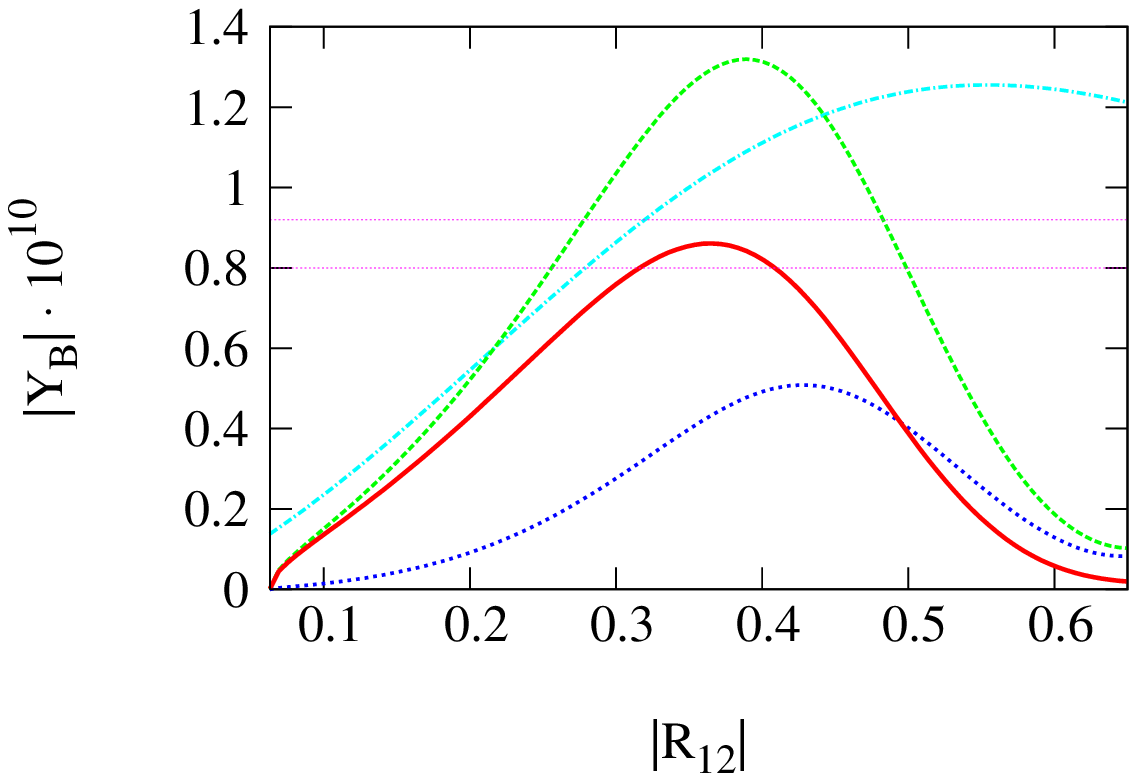}\\
 \includegraphics[width=8.5cm,height=6.5cm]{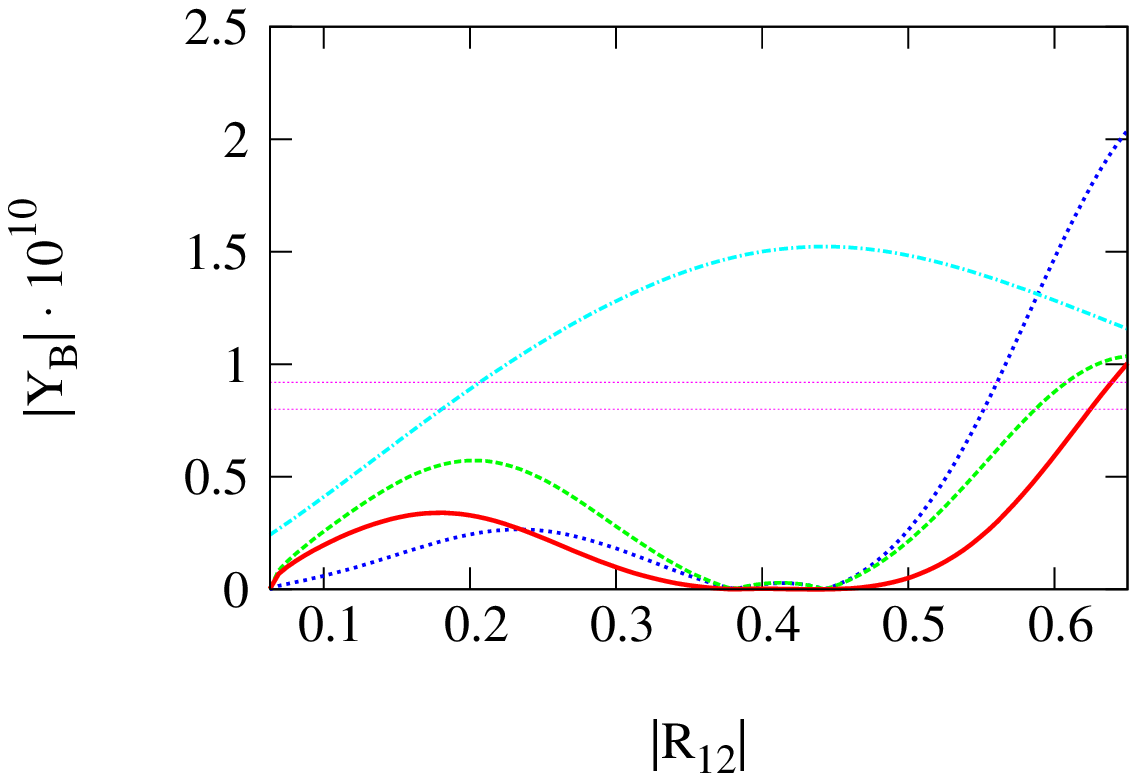}\\
 \includegraphics[width=8.5cm,height=6.5cm]{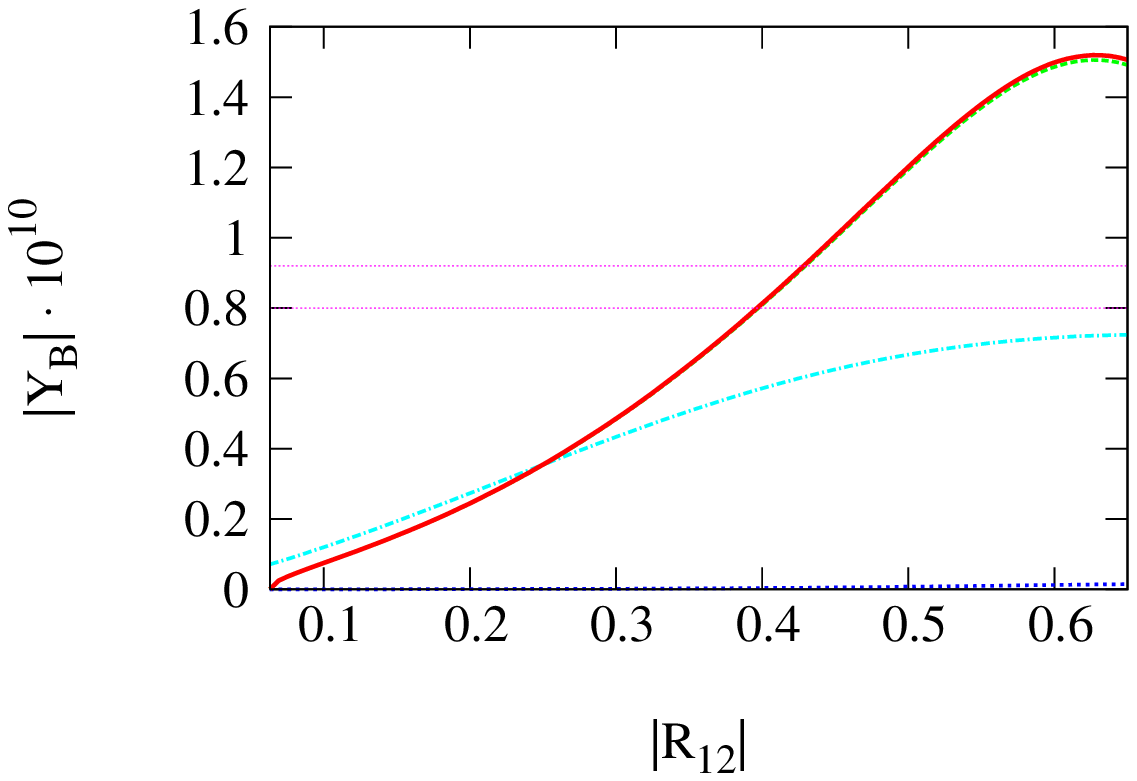}
\end{tabular}
\caption{
\label{IH_YB_a21_0.5b} 
The dependence of 
the ``high energy'' term 
$|Y^0_B A_{\rm HE}|$ (blue line), the ``mixed'' term
$|Y^0_B A_{\rm MIX}|$ (green line) and of the 
total baryon asymmetry $|Y_B|$ (red line) 
on $|R_{12}|$ in the case of IH spectrum, 
CP-violation due to the Majorana phase
$\alpha_{21}$ in $U$ and $R$-phases,
for $\alpha_{21} = \pi/2$, $|R_{11}| \cong 1$,
$M_1 = 10^{11}$ GeV and i) $s_{13}=0$ (upper panel),
ii) $s_{13}=0.2$, $\delta = 0$ (middle panel), 
iii) $s_{13}=0.2$, $\delta = \pi$ (lower panel). 
The light-blue curve 
represents the dependence of $Y_B$ on $|R_{12}|$ for the given
PMNS parameters and CP-conserving matrix R, with 
\cite{PPRio106} $R_{11}R_{12}\equiv ik|R_{11}R_{12}|$, $k=-1$
and $|R_{11}|^2-|R_{12}|^2=1$.
}
\end{center}
\end{figure}

\noindent where $A_{\rm HE(MIX)}\equiv 
C_{\rm HE(MIX)}(\y(0.66\mtil_{\t})-\y(0.71\mtil_2))$, 
$\y(0.66\mtil_{\t})$ and $\y(0.71\mtil_2))$
being the efficiency factors for
\noindent the asymmetries $\eps_\t$ and
$\eps_2$ (see \cite{davidsonetal,davidsonetal2}), 
and 
\begin{eqnarray}
Y^0_B & \cong & 
3\times 10^{-10}\,\left ( \frac{M_1}{10^9~{\rm GeV}}\right )\, 
\left (\frac{\sqrt{\dma}}{5\times 10^{-2}~{\rm eV}}\right )\,,
\label{Y0B}
\end{eqnarray}
\begin{eqnarray}
C_{\rm HE}\, = \, G_{11}\,  
\sin 2\tilde{\f}_{11} 
\left [\,|U_{\t1}|^2-|U_{\t2}|^2\,\right ]\,,
\label{CHE1}
\end{eqnarray}
\begin{eqnarray}
C_{\rm MIX}\, \cong \, 2\, G_{12}\,  
\sin(\tilde{\f}_{11}+\tilde{\f}_{12})\,{\rm Re}(U_{\t1}^*U_{\t2})\,,
\label{CMIX}
\end{eqnarray}
\noindent where $G_{11}\equiv|R_{11}|^2/(|R_{11}|^2+|R_{12}|^2)$, 
$G_{12}\equiv|R_{11}R_{12}|/(|R_{11}|^2+|R_{12}|^2)$ and
we have neglected the contributions proportional to 
the factor $0.5 \dmsol/|\dma|\cong 0.016$ 
in the CP-asymmetry $\eps_\t$.
In Eq. (\ref{YB_2}),
$Y^0_B A_{\rm HE}$ is the ``high energy'' term which
vanishes in the case of a CP-conserving matrix $R$, 
while $Y^0_B A_{\rm MIX}$ is a ``mixed'' term  which,
in contrast to $Y^0_BA_{\rm HE}$, 
does not vanish  when $R$ conserves CP:
it includes the ``low energy'' 
CP-violation, e.g. due to the Majorana phase
$\a_{21}$ in the neutrino mixing matrix.
We recall that the phase $\alpha_{21}$
enters also into the expression 
for the neutrinoless double beta decay 
effective Majorana mass in the case of 
IH light neutrino mass spectrum
\cite{BPP1}. In order to have 
CP-violation due to the  Majorana 
phase $\a_{21}$, both  ${\rm Im}(U_{\t1}^*U_{\t2})$
and ${\rm Re}(U_{\t1}^*U_{\t2})$ should be 
different from zero \cite{JMaj87,ASBranco00}.

 Using the formalism described above,  
we have studied the interplay 
between the CP-violation arising from the 
``high energy'' phases of the orthogonal 
matrix $R$ 
and the ``low energy'' CP-violating Dirac and/or Majorana 
phases in the neutrino mixing matrix, 
as well as the relative contributions of 
the ``high energy'' and the ``mixed'' terms 
$Y^0_B A_{\rm HE}$ and $Y^0_B A_{\rm MIX}$
in $Y_B$. We have found that 
there exist  large regions of the 
corresponding leptogenesis parameter 
space where the ``high energy'' contribution 
to $Y_B$ is subdominant, or even strongly 
suppressed. These results are illustrated 
in Figs. 1 and 2. Below we discuss briefly 
two specific examples of such a suppression  
for $R_{13} = 0$, which can take place even 
when the ``high energy'' $R$-phases 
possess large CP-violating values.
In both cases the asymmetry  $\eps_\t$ is 
produced in the regime of mild wash-out 
($\mtil_{\t} \cong (1 - 3)\times 10^{-3}$ eV), 
while the asymmetry  $\eps_2$ is generated 
with strong wash-out effects (see, e.g. 
\cite{davidsonetal,davidsonetal2}). 
We note also that in both cases 
we have $\eps_2 = -\eps_\t + O(\dmsol/|\dma|)$. 
Under these conditions the two-flavour 
regime in leptogenesis is realised 
typically for
\footnote{We thank A. Riotto for 
useful discussions of this point.}
$M_1 \ltap 5\times 10^{11}$ GeV 
\cite{DiBGRaff06,DeSARio06} 
(see also \cite{DavNN08}).
The are small subregions of the 
parameter space explored by 
us where our results
are valid for  $M_1 \leq 7\times 10^{11}$ GeV;
in another subregion they are
valid for $M_1 \leq 3\times 10^{11}$ GeV.
If, for instance, $|R_{11}| = 1$,
the two-flavour regime of 
leptogenesis is realised for
$M_1 \ltap 5\times 10^{11}$ GeV
provided $|R_{12}| \leq 0.7$.
For $|R_{11}| \leq 0.5$, 
the same conclusion is valid for 
$M_1 \ltap 5\times 10^{11}$ GeV  
in the whole interval of variability of 
$|R_{12}|$;  for $|R_{11}| = 1.1$ 
and $|R_{12}| \leq 1$ this is realised for 
$M_1 \ltap 3\times 10^{11}$ GeV.
In the latter case $|R_{12}|$ can vary in the 
interval $0.45 \ltap |R_{12}| \ltap 1.45$.

  Consider the term $Y^0_B A_{\rm HE}$. 
One can convince oneself that 
for sufficiently large $\theta_{13}$,
$Y^0_B A_{\rm HE}$ depends in a crucial 
way on the Dirac phase $\d$ 
through the following combination of 
the elements of the neutrino mixing matrix:
\begin{eqnarray}
|U_{\t1}|^2-|U_{\t2}|^2 &\cong &(s_{12}^2 - c_{12}^2)s_{23}^2 -
4s_{12}c_{12}s_{23}c_{23}s_{13}\cos\d\nonumber\\
&\cong& -0.20 - 0.92\,s_{13}\,\cos\d\,,
\end{eqnarray}
%
where we have used $s_{12}^2=0.30$ and
$s_{23}^2 = 0.5$. 
Indeed, for, e.g. $s_{13} = 0.2$ and 
the Dirac phase assuming 
the CP-conserving value $\delta = \pi$, 
we get $(|U_{\t1}|^2-|U_{\t2}|^2) \cong (- 0.016)$. 
At the same time we have 
$|Y^0_B A_{\rm MIX}|\propto |U_{\t1}^*\, U_{\t2}| 
\cong 0.27$. As a consequence, 
if the Majorana phase $\a_{21}$ has a 
sufficiently large  CP-violating value,
the contribution of $|Y^0_B A_{\rm MIX}|$ 
to $|Y_B| $ can be by
an order of magnitude bigger than 
the contribution of the ``high energy'' 
term $|Y^0_B A_{\rm HE}|$.
Actually, for $s_{12}^2=0.30$ and
$s_{23}^2 = 0.5$, the ``high energy'' term in
$Y_B$ will be strongly suppressed by 
the factor $(|U_{\t1}|^2-|U_{\t2}|^2)$
if $(-\sin\theta_{13}\cos\d)\gtap 0.15$, 
independently of the values of the 
``high energy'' phases 
$\tilde{\f}_{11}$ and $\tilde{\f}_{12}$. 
Even if the latter assume large CP-violating values, 
the purely ``high energy'' contribution to $Y_B$ 
would play a subdominant role in the generation 
of the baryon asymmetry compatible with the 
observations  if the above inequality holds.
For $(-\sin\theta_{13}\cos\d)> 0.17$ and 
$M_1 \ltap 5\times 10^{11}$ GeV,
the observed value of the baryon asymmetry
cannot be generated by the 
``high energy'' term $Y^0_B A_{\rm HE}$
alone. One can have successful 
leptogenesis in this case only
if there is an additional dominant 
contribution in $Y_B$
due to the CP-violating 
Majorana phase $\a_{21}$ 
in the neutrino mixing matrix.
Let us emphasise that this result is valid 
in the whole range of variability of 
the parameter $|R_{12}|$,
$|(1 -  |R_{11}|^2)| \leq |R_{12}|^2 
\leq (1 +  |R_{11}|^2)$, 
and for $|R_{11}|$ having values 
in the interval $0.3 \ltap |R_{11}| \ltap 1.2$.
For values of $|R_{11}|$ outside the indicated 
interval we cannot have successful leptogenesis 
in the two-flavour regime
for $M_1 \ltap 5\times 10^{11}$ GeV.
For the 3$\sigma$ allowed values of $s_{12}^2=0.38$ 
and $s_{23}^2 = 0.36$, 
the same conclusion is valid if 
$0.06 \lesssim (-\sin\theta_{13}\cos\d) \lesssim 0.12$.
The values of $\sin\theta_{13}$ and 
$\sin\theta_{13}\cos\d$, for which we can have 
the discussed strong suppression 
\footnote{It is interesting to note that 
in the recent analysis of the global neutrino 
oscillation data \cite{Lisis13cd08},
a nonzero value of $\sin^2\theta_{13}$ 
was reported at $1.6\sigma$.  
The best value and the $1\sigma$ 
allowed interval of values of 
$\sin\theta_{13}$ found in 
\cite{Lisis13cd08}, 
$\sin\theta_{13} = 0.126$ and
$\sin\theta_{13} = (0.077 - 0.161)$, 
are in the range of interest for our 
discussion. In addition, 
$\cos\delta = -1$ is reported to be preferred
over $\cos\delta = +1$ by the atmospheric 
neutrino data \cite{Lisis13cd08,Esca08}.
}
of  $Y^0_B A_{\rm HE}$, can be probed by the 
Double CHOOZ and Daya Bay reactor neutrino
experiments \cite{DCHOOZ,DayaB} and by the 
planned accelerator experiments on CP violation 
in neutrino oscillations \cite{Future}.

The results discussed above are illustrated 
in Fig. 1, where we show the 
dependence of $|Y^0_B A_{\rm HE}|$, 
$|Y^0_B A_{\rm MIX}|$ 
and $|Y_B|$ 
on $|R_{12}|$ for 
a fixed value of
$|R_{11}| = 0.7$ ($R_{13}=0$) and 
$\a_{21}=\pi/2$, $s_{13}=0.2$, 
$(-s_{13}\cos\d)= 0.15$ and 
$M_1 = 10^{11}$ GeV.
Note that varying $|R_{12}|$ in its allowed range
is equivalent to varying the ``high energy'' 
CP-violating phases, see eqs. (2) and (3). 
Shown is also the behavior 
of the ``high energy'' 
term  for two additional values of
$(-s_{13}\cos\d)$. 
As is clearly seen in Fig. 1,
for $(-s_{13}\cos\d) \gtap 0.15$,
$|A_{\rm HE}|$ is strongly suppressed 
and is much smaller than 
$|A_{\rm MIX}|$ in almost all 
the range of variability of $|R_{12}|$.
The same conclusion holds 
if we allow $|R_{11}|$ to vary in the 
range $0.3\ltap |R_{11}| \ltap 1.2$.

  Another case in which the contribution of the
``high energy'' term in $Y_B$ is subdominant 
is illustrated in Fig. 2. 
We show the two different contributions 
to the baryon asymmetry as function
of $|R_{12}|$ which is varied in the interval 
$0.05 \leq |R_{12}| \leq 0.65$,
in the case of $|R_{11}| \cong 1$, 
$R_{13}=0$ and 
i) $s_{13}=0$ (upper panel),
ii) $s_{13}=0.2$, $\delta = 0$ (middle panel), 
iii) $s_{13}=0.2$, $\delta = \pi$ (lower panel). 
The behavior of the total baryon asymmetry 
generated when the CP-violation is due 
exclusively to the Majorana
phase $\a_{21}$ \cite{PPRio106} is also given. 
We observe that in most of 
the chosen range of $|R_{12}|$, 
the contribution of the ``mixed'' term 
$|Y^0_B A_{\rm MIX}|$ 
in $|Y_B|$ 
is greater than
that of the ``high energy'' term $|Y^0_B A_{\rm HE}|$ 
and  plays a dominant role 
in the generation of baryon 
asymmetry compatible with that observed. 
Indeed, it follows from eqs. (2) and (3) that 
in the case under discussion we have: 
$\sin 2\tilde{\f}_{11} \cong -\,|R_{12}|^2$ and 
$\sin 2\tilde{\f}_{12} \cong (1 - |R_{12}|^4/8)$.
This implies 
$|A_{\rm HE}| \propto |G_{11}\sin 2\tilde{\f}_{11}| 
\propto |R_{11}R_{12}|^2$, while
$|A_{\rm MIX}| \propto 
2|G_{12}\sin (\tilde{\f}_{11} + \tilde{\f}_{12})| 
\propto \sqrt{2}|R_{11}R_{12}|$, where we have used 
$|\sin (\tilde{\f}_{11} + \tilde{\f}_{12})| \cong 1/\sqrt{2}$.
The latter approximation is rather accurate for 
$|R_{11}| = 1$ and $|R_{12}|\leq 0.5$.
Thus, for $|R_{12}| = 0.4$, for instance, we have
$\tilde{\f}_{11} \cong -0.08$, $\tilde{\f}_{12} \cong \pi/4$,
and correspondingly for $s_{13} = 0$ and 
$\alpha_{21} = \pi/2$ we get
$|A_{\rm MIX}|/|A_{\rm HE}| \cong 2.6$ 
(Fig. 2, upper panel).
Note also that, e.g. in the case of 
$s_{13}=0$, the generated $|Y_B|$ 
is largest when the ``high energy'' 
$R$-phases assume CP-conserving values.
The same feature is clearly observed also
for $s_{13}=0.2$ and $\delta = 0$
at $|R_{12}|\ltap 0.55$.
Moreover, for  $0.25\ltap |R_{12}| \ltap 0.50$, 
the baryon asymmetry generated in the case of 
CP-conserving $R$-phases is significantly larger 
in absolute value than the asymmetry produced 
when the relevant $R$-phases possess 
CP-violating values (Fig. 2, middle panel).
We note also that for $s_{13}=0.2$ and $\delta = \pi$
(Fig. 2, lower panel), the ``high energy'' term 
$|Y^0_B A_{\rm HE}|$ is strongly suppressed 
by the factor $(|U_{\t1}|^2-|U_{\t2}|^2)$ 
(see eq. (11) and the related discussion) 
and is hardly visible in the corresponding 
figure. If, however, $|R_{12}| \gtap 0.8$ 
and $M_1 \gtap 7\times 10^{10}$ GeV,
the ``high energy'' term in $|Y_B|$
is the dominant one and can provide 
the requisite baryon asymmetry 
compatible with the observations.

  In summary, we have considered a 
case of subdominant/suppressed 
``high energy'' contribution to the baryon asymmetry 
of the Universe in thermal flavoured leptogenesis 
scenario with hierarchical heavy Majorana 
neutrinos. It can arise if the
light neutrino mass spectrum is of
inverted hierarchical type and if 
the $R_{13}$ element of the complex 
orthogonal $R$-matrix 
in the orthogonal parametrisation 
of neutrino Yukawa couplings 
satisfies ${\rm Im}(R^2_{13}) = 0$.
Under these conditions the 
``high energy'' contribution to 
the baryon asymmetry, 
arising in the two-flavour regime 
in leptogenesis,
will be strongly suppressed 
by the factor $(|U_{\t1}|^2-|U_{\t2}|^2)$  
provided $(-\sin\theta_{13}\cos\d) \gtap 0.15$. 
The interval of values of $(-\sin\theta_{13}\cos\d)$ 
for which the suppression takes place
depends on the precise values of
$\sin^2\theta_{12}$ and $\sin^2\theta_{23}$, 
which are still determined experimentally with  
non-negligible uncertainty.
In the simplified case of $R_{13} = 0$,
this result is valid 
in the whole range of variability of 
the parameter $|R_{12}|$, determined by 
\cite{Molinaro:2008rg} 
$|(1 -  |R_{11}|^2)| \leq |R_{12}|^2 
\leq (1 +  |R_{11}|^2)$, and
for  $|R_{11}|$ having a value in the interval 
$0.3 \ltap |R_{11}| \ltap 1.2$.
For the indicated ranges of values of 
$|R_{11}|$ and $|R_{12}|$, the 
``high energy'' CP-violating phases 
are not necessarily small.
However, reproducing the observed value of the
baryon asymmetry is problematic (or can even be 
impossible) without 
a contribution due to the CP-violating 
phases in the PMNS matrix.
The ``high energy''contribution can be 
subdominant also in the case of 
$\sin\theta_{13}= 0$. This possibility 
can be realised 
for values of the Majorana phase 
in the PMNS matrix 
$0 < \alpha_{21} \ltap 2\pi/3$  and 
roughly in half of the 
parameter space spanned by the relevant 
elements of the $R$ matrix. In both cases 
the observed value of the baryon asymmetry 
can be reproduced for values of the lightest
RH Majorana neutrino mass lying in the interval
$5\times 10^{10}~{\rm GeV} \ltap M_1 \ltap 7\times 10^{11}$ GeV.
Similar results hold in the more general case of 
$R_{13}\neq 0$, ${\rm Im}(R^2_{13}) = 0$, for 
$0 \leq |R_{13}|\ltap 0.9$,
$1.05 \ltap |R_{13}|\ltap 1.5$, and
$0.3 \ltap |R_{11}|\ltap 1.2$.

  The results obtained in this  
study show that in the ``flavoured'' 
leptogenesis scenario, the contribution 
to $Y_B$ due to the ``low energy'' CP-violating 
Majorana and Dirac phases in the neutrino 
mixing matrix, in certain physically interesting cases, 
like IH light neutrino mass spectrum, 
relatively large value of $(-\sin\theta_{13}\cos\d)$, etc., 
can be indispensable for the generation of 
the observed baryon asymmetry of the Universe 
even in the presence of ``high energy'' 
CP-violation, generated by additional physical 
phases in the matrix of neutrino 
Yukawa couplings, e.g. by CP-violating phases in 
the complex orthogonal matrix $R$ appearing in 
the ``orthogonal parametrisation'' 
of neutrino Yukawa couplings.

 {\bf Acknowledgments.} This work was supported 
in part by the Italian INFN and MIUR under 
the programs ``Fisica Astroparticellare''   
and  ``Fundamental Constituents of the Universe'', 
and by the EU Network ``UniverseNet'' 
(MRTN-CT-2006-035863).

\section*{ADDENDUM}

We have shown in the article that 
for a certain range of values of 
$\sin(\theta_{13})\cos(\delta)$ 
in the case of IH light neutrino mass spectrum and 
hierarchical heavy Majorana neutrino mass spectrum, 
the ``high energy'' contribution to the baryon 
asymmetry of the Universe will be strongly suppressed 
if the complex $R_{13}$ element of the $R$-matrix 
satisfies the inequality:  
$|R_{13}|^{2}|\sin(2\tilde{\varphi}_{13})|\ll \text{min}(\,|R_{11}|^{2}|\sin(2\tilde{\varphi}_{11})|,|R_{12}|^{2}|\sin(2\tilde{\varphi}_{12})|\,)$. 
In this Addendum we extend the analysis performed in our article 
to the case of arbitrary complex $R_{13}$. 
Hence, we abandon the condition on  $|R_{13}|^{2}|\sin(2\tilde{\varphi}_{13})|$ 
given above and determine the ranges of values of 
$|R_{13}|$ for which the indicated strong 
suppression of the ``high energy'' contribution 
to the baryon asymmetry still takes place 
even when the phase of $R_{13}$, $\tilde{\varphi}_{13}$, 
is allowed to assume large CP-violating values. 
In this case the contribution to the corresponding 
CP asymmetry due to the low energy 
Majorana CP violating 
phases is essential for producing baryon asymmetry 
compatible with the observations.

   It follows from the numerical analysis 
performed in this Addendum that for, 
e.g., $M_{1}=10^{11}$ GeV, the scenario 
described above is realised 
and we have subdominant (strongly suppressed) 
``high energy'' contribution 
to the observed value of the baryon asymmetry  
for arbitrary  $\tilde{\varphi}_{13}$ if
$i)$ for $|R_{11}|<0.5$,
$|R_{13}|$ satisfies
$|R_{13}|\lesssim|R_{11}|$;
$ii)$ for $0.5\lesssim |R_{11}|<1$ we have 
$|R_{13}|<0.5$, and if 
$iii)$ for $|R_{11}|>1$ we have
$|R_{13}|<|R_{11}|/2$. 
In each of the cases $i)$ - $iii)$
we can have successful 
leptogenesis, as our results show,
due to the contribution 
to the baryon asymmetry associated with 
the Majorana CP violating phase(s) in 
the neutrino mixing matrix.

\begin{figure}[h]
\begin{center}
\begin{tabular}{cc}
\includegraphics[width=6.5truecm,height=5.5cm]
{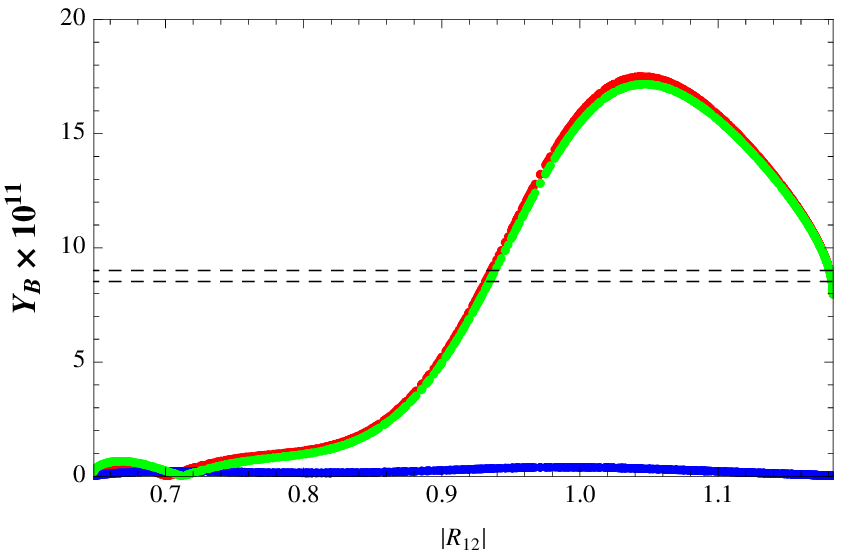}&
\includegraphics[width=6.5truecm,height=5.5cm]
{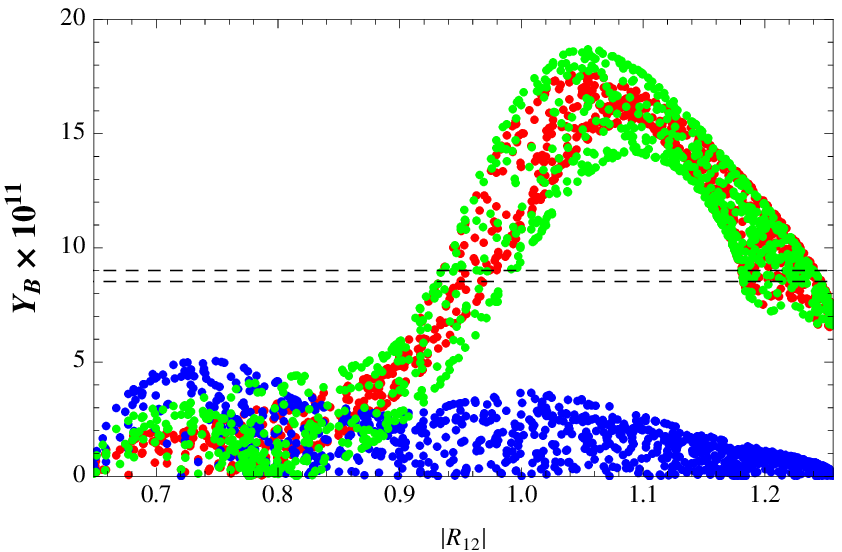}\\
\includegraphics[width=6.5truecm,height=5.5cm]
{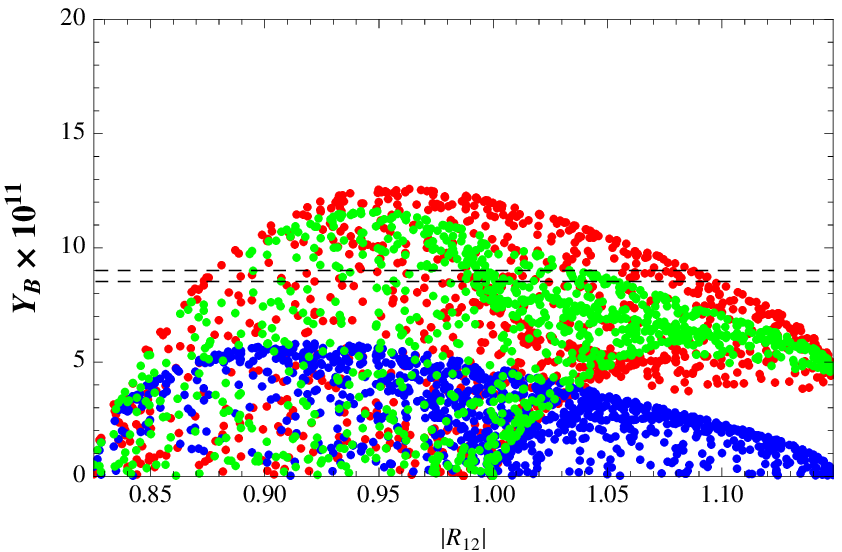}&
\includegraphics[width=6.5truecm,height=5.5cm]
{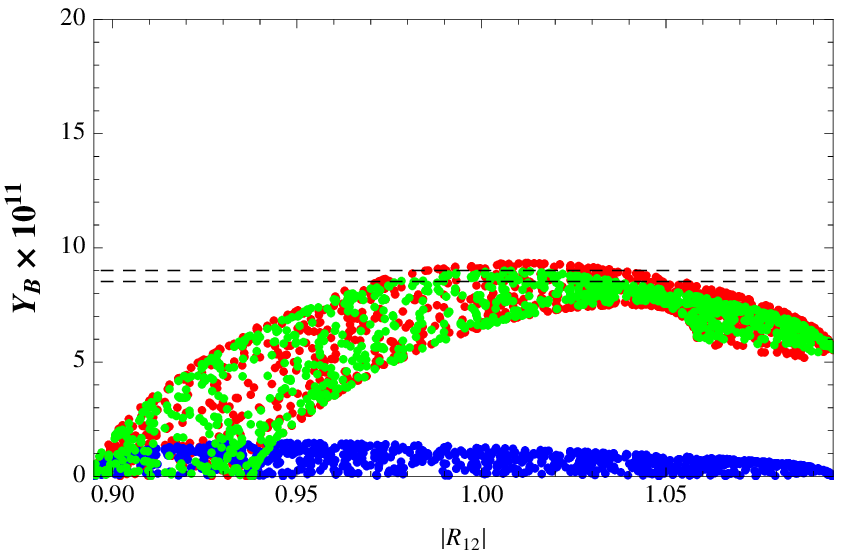}\\
\includegraphics[width=6.5truecm,height=5.5cm]
{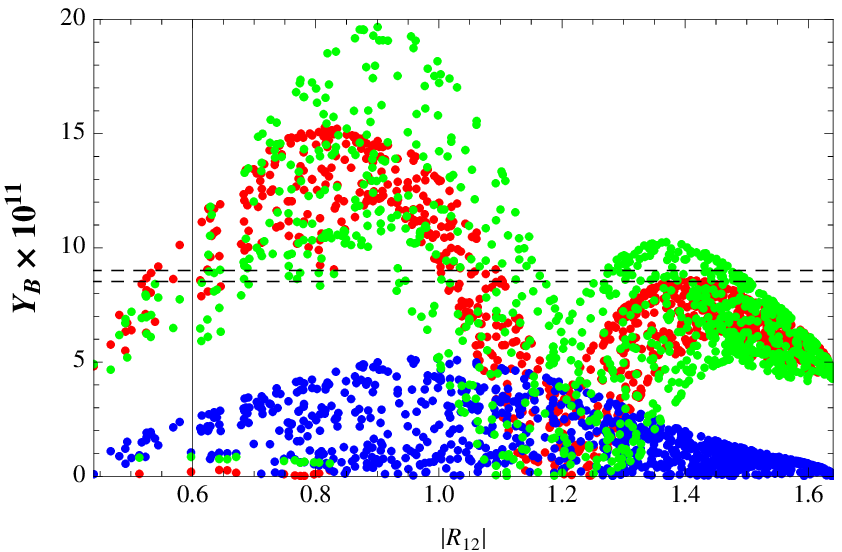}&
\includegraphics[width=6.5truecm,height=5.5cm]
{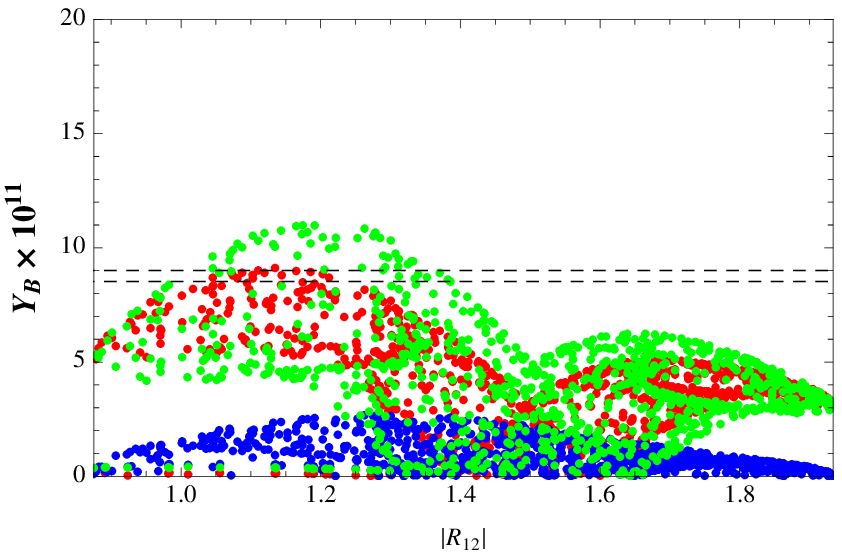}
\end{tabular}
\caption{\label{Fig3}Total baryon asymmetry $Y_{B}$ (red dots), 
``high energy'' term (blue dots) and
``mixed'' term (green dots) vs $|R_{12}|$ for 
$M_{1}=10^{11}$ GeV, $m_{3}=0$, $\alpha_{21}=\pi/2$, $s_{13}=0.2$, 
$\delta=\pi$, complex $R_{11}$, $R_{12}$,
and $i)$ $|R_{11}|=0.7$ and  real $R_{13}=0.3$ (upper left panel); 
$ii)$ $|R_{11}|=0.7$ and $|R_{13}|=0.3$ (upper right panel);
$iii)$ $|R_{11}|=0.4$ and $|R_{13}|=0.4$ (middle left panel);
$iv)$ $|R_{11}|=0.4$ and 
$|R_{13}|=0.2$ (middle right panel);
$v)$ $|R_{11}|=1.3$ and $|R_{13}|=0.6$ (lower left panel);
$vi)$ $|R_{11}|=1.5$ and $|R_{13}|=0.7$ (lower right panel).
The figures in the upper right panel, the middle and lower panels
are obtained for complex $R_{13}$.
The horizontal lines indicate the allowed range 
of  the baryon asymmetry: $Y_{B}=(8.77\pm0.24)\times10^{-11}$. 
}
\end{center}
\end{figure}

 The results obtained in the present Addendum 
are illustrated in Fig.~\ref{Fig3}, 
where we show the total baryon asymmetry $Y_{B}$ (red dots/line), the 
purely ``high energy'' contribution (blue dots/line) and
the ``mixed'' term as function of $|R_{12}|$ for 
fixed values of $|R_{11}|$ and $|R_{13}|$. 
The lightest neutrino mass $m_{3}$ 
is set to zero and 
the lightest RH Majorana neutrino mass 
corresponds to $M_{1}=10^{11}$ GeV. 
The figure is obtained for $\sin(\theta_{13})\cos(\delta)=-0.2$.
In the upper left panel, we have set 
$|R_{11}|=0.7$ and have taken a real $R_{13}$, 
with $R_{13}=0.3$. The phase of $R_{11}$, $\tilde{\varphi}_{11}$, 
was varied in the interval $[3\pi/2,2\pi]$, while the 
$R_{12}=|R_{12}|e^{i\tilde{\varphi}_{12}}$ is determined by the orthogonal
condition: $R_{11}^{2}+R_{12}^{2}+R_{13}^{2}=1$. 
In all the other panels of 
Fig.~1 we consider a complex  $R_{13}=|R_{13}|e^{i\tilde{\varphi}_{13}}$ 
and the ``high energy'' phase $\tilde{\varphi}_{13}$ 
was assumed to take random values in the interval $[0,2\pi]$. 
We plot  the behavior of each term contributing 
to the baryon asymmetry as a function of $|R_{12}|$ 
for fixed $|R_{11}|$ and $|R_{13}|$, following the procedure 
described above.

\end{document}